\documentclass{article}
\usepackage{spconf,amsmath,graphicx}
\usepackage[LGR, T1]{fontenc}
\usepackage[utf8]{inputenc}
\usepackage[greek,english]{babel}
\usepackage{alphabeta}
\usepackage{hyperref}

\title{Evaluating the Intelligibility Benefits of Neural Speech Enrichment for Listeners with Normal Hearing and Hearing Impairment using the Greek Harvard Corpus}
\name{ Muhammed PV Shifas$^1$, Anna Sfakianaki$^1$, Theognosia Chimona$^2$, Yannis Stylianou$^1$}
\address{
  $^1$Computer Science Department, University of Crete, Greece\\
  $^2$ENT Department, Chania General Hospital, Greece}
\begin{document}
%
\maketitle
\begin{abstract}
In this work we evaluate a neural based speech intelligibility booster based on spectral shaping and dynamic range compression (SSDRC), referred to as WaveNet-based SSDRC (wSSDRC), using a recently designed Greek Harvard-style corpus. The corpus has been developed according to the format of the Harvard/IEEE sentences and offers the opportunity to apply neural speech enhancement models and examine their performance gain for Greek listeners.
wSSDRC has been successfully tested for English material and speakers in the past. In this paper we revisit wSSDRC to perform a full scale evaluation of the model with Greek listeners under the condition of equal energy before and after modification. Both normal hearing (NH) and hearing impaired (HI) listeners evaluated the model under speech shaped noise (SSN) at listener-specific SNRs matching their Speech Reception Threshold (SRT) - a point at which 50 \% of unmodified speech is intelligible. The analysis statistics show that the wSSDRC model has produced a median intelligibility boost of 39\% for NH and 38\% for HI, relative to the plain unprocessed speech.
\end{abstract}
\begin{keywords}
speech intelligibility, neural network approach, subjective intelligibility evaluation, Greek Harvard corpus.
\end{keywords}
\section{Introduction}
\label{sec:intro}

Speech produced in everyday life might suffer from low intelligibility originating from masking effects due to environmental noise. Understanding noise-masked speech requires a certain amount of cognitive effort, depending on the noise level~\cite{simantiraki2018impact}. To mitigate the masking effects, the speaker tries to adjust speech production, often involuntarily -- which is known as Lombard effect~\cite{junqua1996influence}. When compared to plain speech recorded in quiet, Lombard speech is generally characterized by higher $F0$ and format frequencies, and greater energy in the mid-high frequency bands, resulting in reduced spectral tilt~\cite{stanton1988acoustic}. Lombard speech is more intelligible than plain speech under similar amounts of noise masking~\cite{lu2009contribution}\cite{cooke2012intelligibility}\cite{cooke2014contribution}.
\vspace{.3cm}

Parallel to the advancements in phonetic research, speech engineers started developing signal processing models that aim at improving listening comfort in noise. High-pass filtering as well as sharpening of formant regions has been shown to help understand speech better, while dynamic range compression, a popular technique in audio engineering, has been widely used in modification strategies~\cite{zorila2012speech}\cite{schepker2015speech}. Over the years, this task has been addressed mainly by signal processing approaches, due to their inherent simplicity~\cite{cooke2013intelligibility}. However, the fundamental problem with these approaches is that they are highly sensitive to noise; if there is noise at the recording end  -- something very common in practical scenarios -- the system performance is dramatically degraded. This has not been often noted in the literature, as most of these systems are designed and tested on ideal conditions, such as in a sound-proof, isolated booth. To enable its smooth operation in outdoor environments, one must employ a noise reduction front-end module to prevent propagation of recording noise into the intelligibility enhancer ~\cite{griffin2015improved}\cite{khademi2017intelligibility}. Such an approach would not be optimal in practice as the noise reduction module could introduce additional artifacts, which results to uneven modifications~\cite{zorila2017quality}.

A neural processing perspective of the problem would give higher degree of freedom, since neural models have been proved to be more robust, compared to the pure signal processing approaches, against recording noise \cite{shifas2020fully}\cite{muhammed2019non}. Besides, having a neural speech enrichment module would ease the effort to integrate the intelligibility factor into neural-based speaking devices, like advanced text-to-speech (TTS) systems. Motivated by these observations, we recently proposed a neural intelligibility enhancement model referred to as wSSDRC~\cite{muhammed2018speech}. Its architecture resembles that of the well-known WaveNet model~\cite{oord2016wavenet}, while the model is trained to mimic the modifications performed by another recently proposed algorithm, the spectral shaping and dynamic range compression (SSDRC) algorithm~\cite{zorila2012speech}. An initial evaluation of the model has been conducted in~\cite{pvbenefits}. However, unlike signal processing techniques, neural networks are data-driven and would be sensitive to the data set on which they are trained. Equally, the features learned depend on the linguistic characteristics of the corpus being used to train the model.

Because of data scarcity, we have not been able to test the effectiveness of the model in full scale until now. The creation of a novel, Greek Harvard-style corpus (GrHarvard Corpus) provided the opportunity to revisit the wSSDRC model with Greek speech data. A full scale training of the model on the Greek language has been performed using the majority of the GrHarvard corpus samples, followed by a wider testing on the remaining sentences. Listeners with both normal hearing and hearing impairment have been recruited in the study. Since Greek has significant differences from latin-originated languages which have been broadly used in intelligibility experiments, it is interesting to see how the model learns the modification task. 

The rest of the paper is organized as follows. In Section~\ref{Section.2}, we provide an overview of the wSSDRC model. The details about the GrHarvard corpus are included in Section~\ref{Section.3}. Section~\ref{Section.4} provides information about the listener groups, followed by the results and discussion in Section~\ref{Section.5}. Section~\ref{Section.6} concludes the paper. 

\section{WaveNet-Based Spectral Shaping and Dynamic Range Compression (\lowercase{w}SSDRC)}
\label{Section.2}
WaveNet was initially suggested as a generative model to synthesize high quality speech from text, accumulating phonetic and acoustic representations. The model operates on the waveform domain taking raw speech samples as input and generating the samples in an autoregressive manner. The network architecture of wSSDRC differs from WaveNet in the sense that the autoregressiveness has been omitted to make the generation process faster while keeping the same high quality performance. The resulting model is a regression model that could generate the entire speech segment in a single shot. 

\begin{figure}[h]
  \centering
  \includegraphics[width=\linewidth, height=5cm]{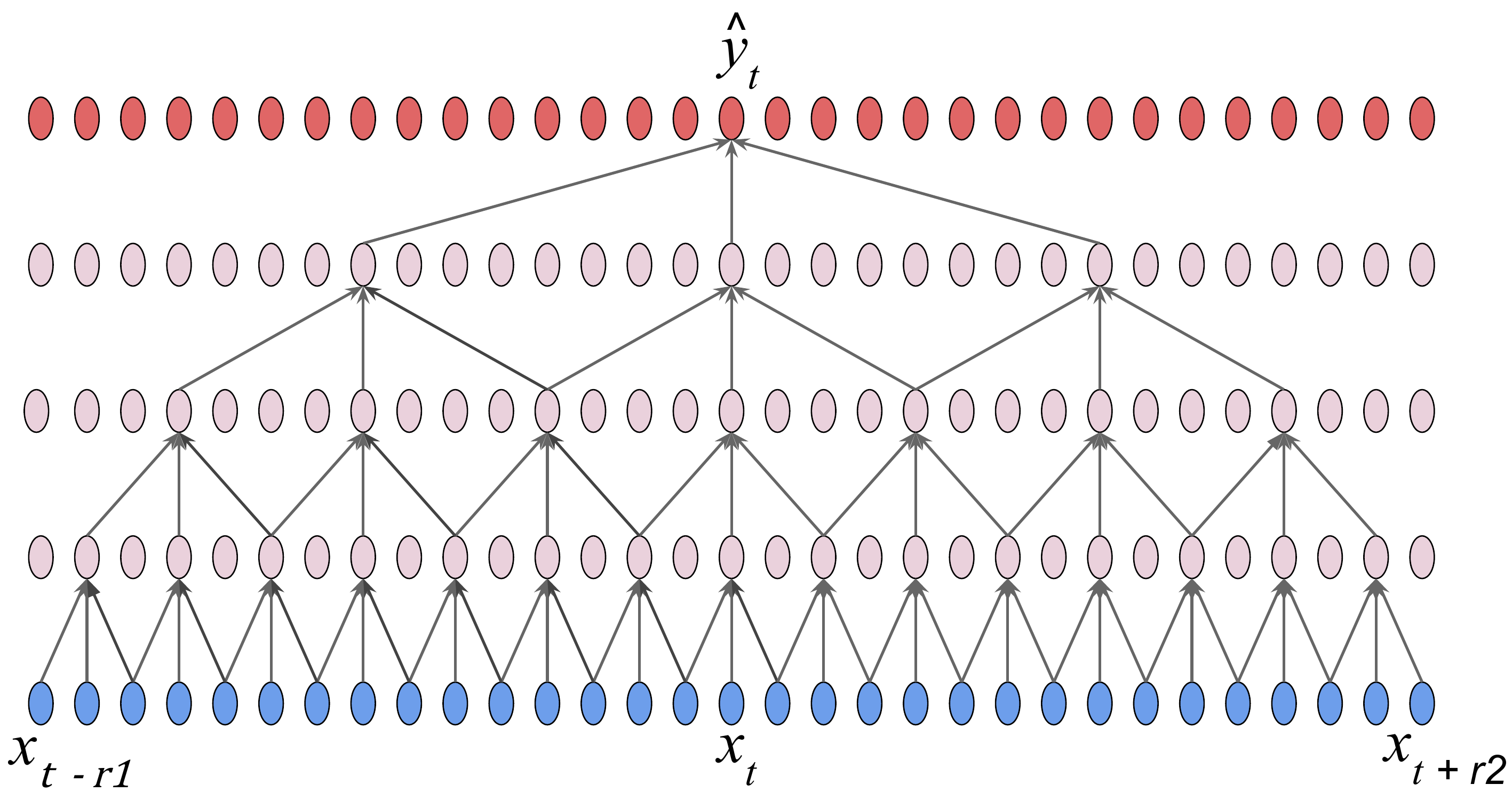}
  \caption{Dilation pattern of the wSSDRC architecture.}
  \label{fig:wavenet}
\end{figure}

The speech modification problem is postulated as a regression task, where the model's objective is to take plain speech samples at the input, $x_t$, and modify its characteristics to create a more intelligible output, $\hat{y}_t$. The entire modification induced by the model can be mathematically stated as
\begin{equation}
\hat{y}_t = \hat{f}(x_{t-r1}, \ldots, x_{t-1}, x_t, x_{t+1}, \ldots, x_{t+r2};  \Theta)
\label{eq1}
\end{equation} 
where $\Theta$ is the model parameters, that need to be optimised for the task. The conditional dependency of the past, $x_{t-r1}$, and future, $x_{t+r2}$, input samples is achieved through the dilated convolution architecture of the model, as shown in Figure~\ref{fig:wavenet}. The network can be causal or non-causal depending on whether to consider ($r2\neq0$) or not consider ($r2=0$) future samples when designing the model architecture. The wSSDRC model follows the non-causal architecture, with $r1=r2=r$. During training, the parameters $\Theta$ are optimized such that the learning function $\hat{f}$ lies as close as possible in the vicinity of the actual function $f$, i.e., $\hat{f}\approx f$.

The main concern is about the target function ($f$) and the kind of modification the network is expected to learn, which reflects the articulatory style modification to be mimicked by the network. One could set the model to mimic natural intelligibility modifications, like in Lombard speech, as long as they satisfy the time alignment constraint stated in (~\ref{eq1}). However, since multiple studies have shown considerable intelligibility gain of SSDRC-processed speech over Lombard speech~\cite{cooke2013evaluating}\cite{cooke2013intelligibility}, we decided that the model should learn an SSDRC-style modification. 

\begin{figure}[h]
  \centering
  \includegraphics[width=\linewidth, height=5.5cm]{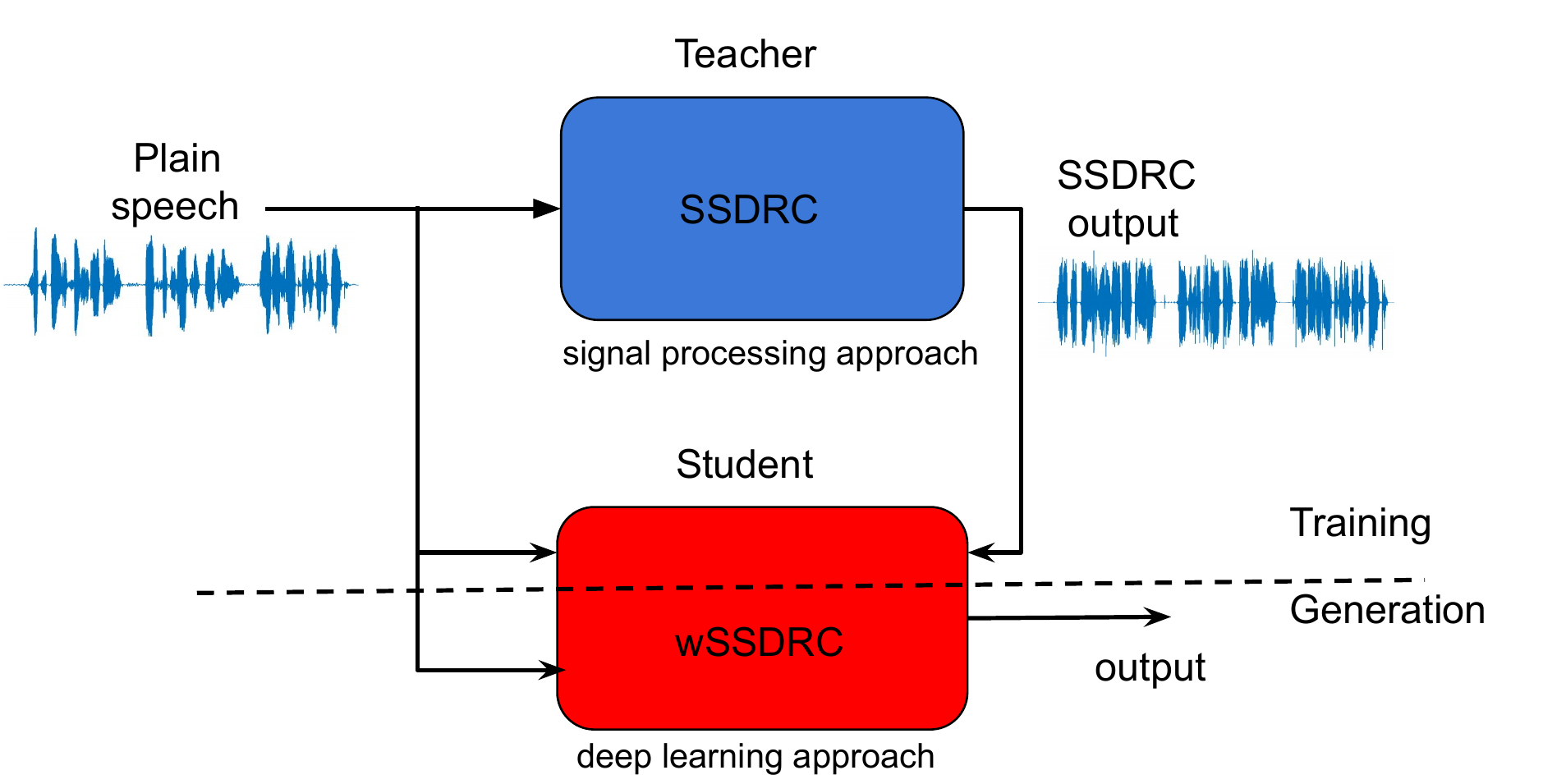}
  \caption{Teacher-Student framework followed to train the wSSSDRC model.}
  \label{fig:teacher-student}
\end{figure}

This has been accomplished by setting the SSDRC (signal processing approach) as the teacher-network to expose the neural model (wSSDRC) to the modification patterns to be learned. Figure~\ref{fig:teacher-student} depicts the aforementioned teacher-student framework.

Since the model is operating in an end-to-end fashion on the waveform domain, the deviation of the prediction from the target is calculated as the average absolute difference between the predicted sample $\hat{y}_t$ and the target sample ${y}_t$ . For an input-target wave pair ($x^{(k)}$,$ y^{(k)}$), the loss function is computed as
\vspace*{.6cm}
\begin{equation}
L(x^{(k)}, y^{(k)}) = \frac{1}{T^{(k)} - 2r} \sum_{t=r}^{T^{(k)}-r}| y^{(k)}_t - \hat{y}^{(k)}_t | 
  \label{eq2}
\end{equation}
where $T^{(k)}$ is the length of signals $x^{(k)}$ and $y^{(k)}$.

\section{The Greek Harvard Corpus}
\label{Section.3}
The Greek Harvard (GrHarvard) Corpus was recently designed to address a lack of Greek sentence corpora developed for intelligibility testing~\cite{Sfakianaki}. It comprises 720 sentences in the format of the Harvard/IEEE material~\cite{rothauser1969ieee} with the necessary accommodations for the Greek language. The original Harvard material has been used extensively in speech intelligibility experiments (e.g. \cite{cooke2013evaluating}, \cite{hu2010importance}) and has also been adapted for the Spanish language~\cite{aubanel2014sharvard}. Each sentence of the GrHarvard Corpus includes five keywords consisting of one, two or three syllables, with the total number of words per sentence varying strictly from five to nine. Sentence content was inspired in part by the original Harvard sentences; a translation of the original material was not possible in most cases, because grammatical differences between the English and the Greek language rendered many of the keywords unsuitable candidates for the GrHarvard Corpus. The majority of keywords have been selected from GreekLex 2~\cite{kyparissiadis2017greeklex}  so that the resulting sentences are meaningful, semi-predictable and resemble everyday language. For example, 
\textgreek{"Το ξύλο είναι άριστο υλικό για παιχνίδια και κύβους"} [to."ksilo."ine."aristo.ili"ko.jja.pe"xniDja.ce."civus] (Wood is an excellent material for toys and cubes), 
\textgreek{“Καυτός ατμός ξέφυγε από τη σπασμένη βαλβίδα”} [ka"ftos.a"tmos."ksefijje.a "po.ti.spa"zmeni.val"viDa]
(Hot steam escaped from the broken valve). The GrHarvard Corpus is freely available to the research community for non-commercial purposes. The 720 sentences in Greek orthography and phonetic transcription as well as metadata information are provided
\footnote{\url{https://www.csd.uoc.gr/~asfakianaki/GrH.html}}.

\section{Experimental Setup}
\label{Section.4}
The 720 utterances of the GrHarvard Corpus were divided into two groups, 600 for training and the remaining 120 for validating and testing the model. We used the same samples as the validation and test set. Sentences with a maximum of 7 words in total were selected for testing / validating. Although the dataset was recorded at $44.1$ kHz, it was downsampled to $16$ kHz, as feeding high-resolution samples into the model would limit the phoneme context covered by the receptive fields. The corresponding target pairs were generated by running the SSDRC algorithm over the samples. 

\textbf{The model specification:} The wSSDRC model has in total 30 layers made up by thrice repeating a block of depth 10 that has the dilation factors [1, 2, 4, 8, 16, 32, 64, 128, 256, 512], starting from the beginning. It sums up to a receptive field of size 6138 ( 3069 past \& 3069 future samples), which means it considered 0.38 s of input samples (for 16 kHz signal) when predicting a single clean sample.  In all the layers, convolutions of 256 channels are used. During training, the target samples predicted in a single traverse is a set of 4096 (training target field size). The model is fed with a single data point every time with a batch size of 1. In the testing phase, the target field size being varied depends on the test frame length. Just before feeding into the model, the wave files have been normalized to an RMS level of 0.06. This removed the loudness variations among the wave files.
The loss function in~(\ref{eq2}) was optimized with the Adam optimization algorithm, with an exponential decaying learning rate method. The hyper parameters of the exponential decay method are -- learning rate = $0.001$, decay steps = $20000$, and decay rate = $0.99$.

In the process of finding the optimal configuration, the model trained with British English was tested on the Greek test set. It performed well, except for some occasional clicks in the generated samples that would make listening less comfortable. Therefore, the Greek training set was ultimately selected to fully train the network. As such, the final evaluating model is purely trained on the Greek Harvard corpus.

Since the primary objective of our work is to measure the wSSDRC modification benefits and compare them to the SSDRC approach that has been used to train the model, wSSDRC and SSDRC are the main models to be evaluated here. Plain speech is also included as a baseline on which the relative intelligibility gains are observed. Experiments have been conducted under the equal total energy constrain: that is, the sentence level energy of modified speech (by SSDRC or wSSDRC) should remain the same as that of plain speech. 

\subsection{Listening Groups}

In order to evaluate the intelligibility of samples generated by wSSDRC, a detailed subjective evaluation was carried out. The evaluation is based on the hypothesis that the wSSDRC model should generate equally intelligible samples compared to the SSDRC. 
Two groups of listeners were recruited: individuals with normal hearing (NH) and hearing impairment (HI). The participants with HI were screened for hearing loss via Pure Tone Audiometry (PTA) at frequencies of $0.5$, $1$, $2$, $4$ kHz in both ears. The group with HI was characterized by an average hearing loss of 62 dBHL. Most of the participants wore hearing aids which were removed during the test.

After examining the participants' responses to the test, four NH participants were excluded due to biased listening. Hence, in the final evaluation $13$ participants with NH and $11$ with HI were included.

\subsection{Masking Noise}
The current evaluation has considered masking based on stationary speech shaped noise (SSN) only. SSN was selected from the Hurricane Challenge~\cite{cooke2013intelligibility}. Since intelligibility level varies from subject to subject, intelligibility gains should be observed from a common reference point.
This was achieved by designing subject-specific Signal-to-Noise Ratio (SNR) sets to match the speech reception threshold (SRT), i.e. the point at which $50$\% of speech is intelligible for each individual listener.

For this, an initial pilot study was carried out, during which each participant was asked to listen to an initial set of samples, masked with SSN at SNR points in the range of $-7$ dB to $-1$ dB for NH and $-3$ dB to $+9$ dB for HI individuals. After analysing the responses, subject-specific SNRs were selected that matched each listener's SRT. The masking noise level for the final test was set on this SNR value.

\vspace{.3cm}
 The speech samples from different models were Root-Mean-Square (RMS) normalized to a level of -23dB before being masked by the noise. The noise segments were scaled by the fraction matching the listener's SNR level, without touching the normalized speech. Since participants listening at their SRT points has been ensured through the pilot study, the intelligibility ceiling due to listener variability was not a factor of concern. In each condition, the participants had to listen to $8$ sentences, with $5$ keywords each and a total word count not exceeding $7$ per sentence. HI listeners have participated in a controlled environment (ENT clinic at the hospital). NH listeners, however, had to participate from their homes because of the current locked-down situation. NH have been instructed to use the same listening conditions between the pilot and the main listening test (speakers, headphones etc). Each sentence has been heard only once by the participants. 

\section{Results and Discussion}
\label{Section.5}
\begin{figure}[h]
  \centering
  \includegraphics[width=\linewidth, height=6cm]{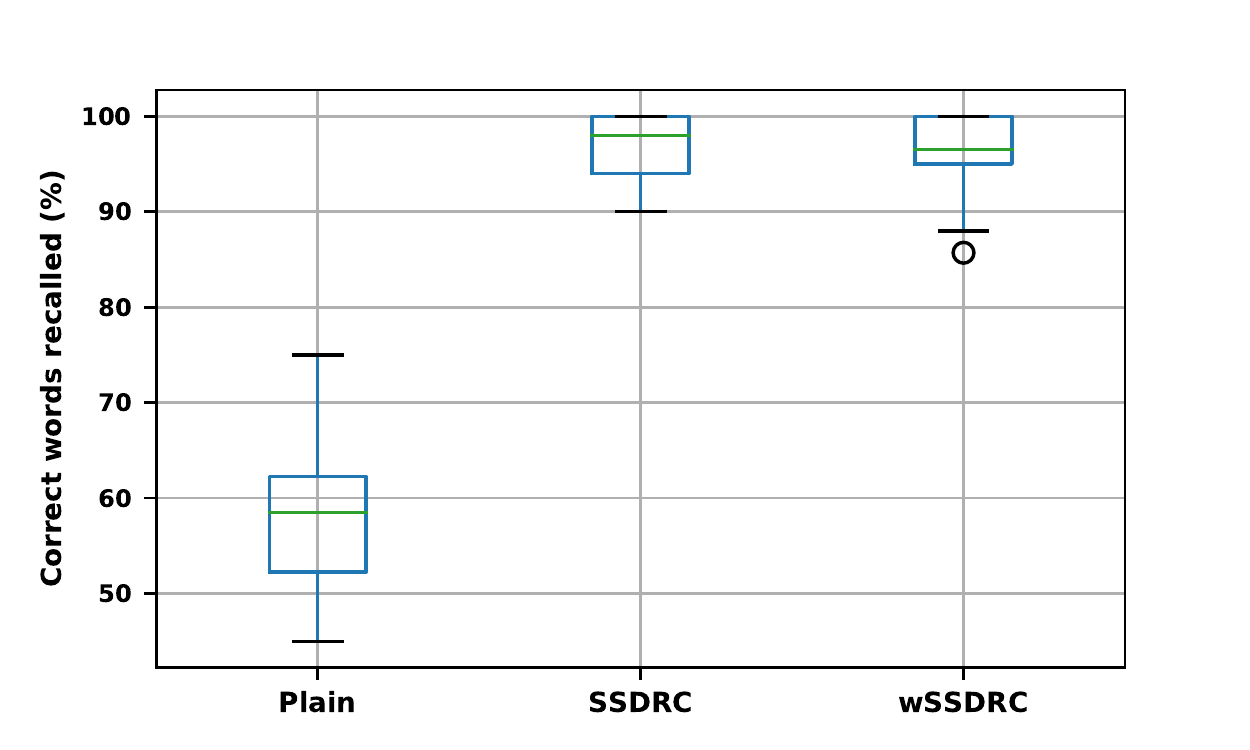}
  \caption{Words recalled by participants with Normal Hearing (NH) in different conditions; boxes represent data dispersion}
  \label{fig:NH}
\end{figure}

\begin{figure}[h]
  \centering
  \includegraphics[width=\linewidth, height=6cm]{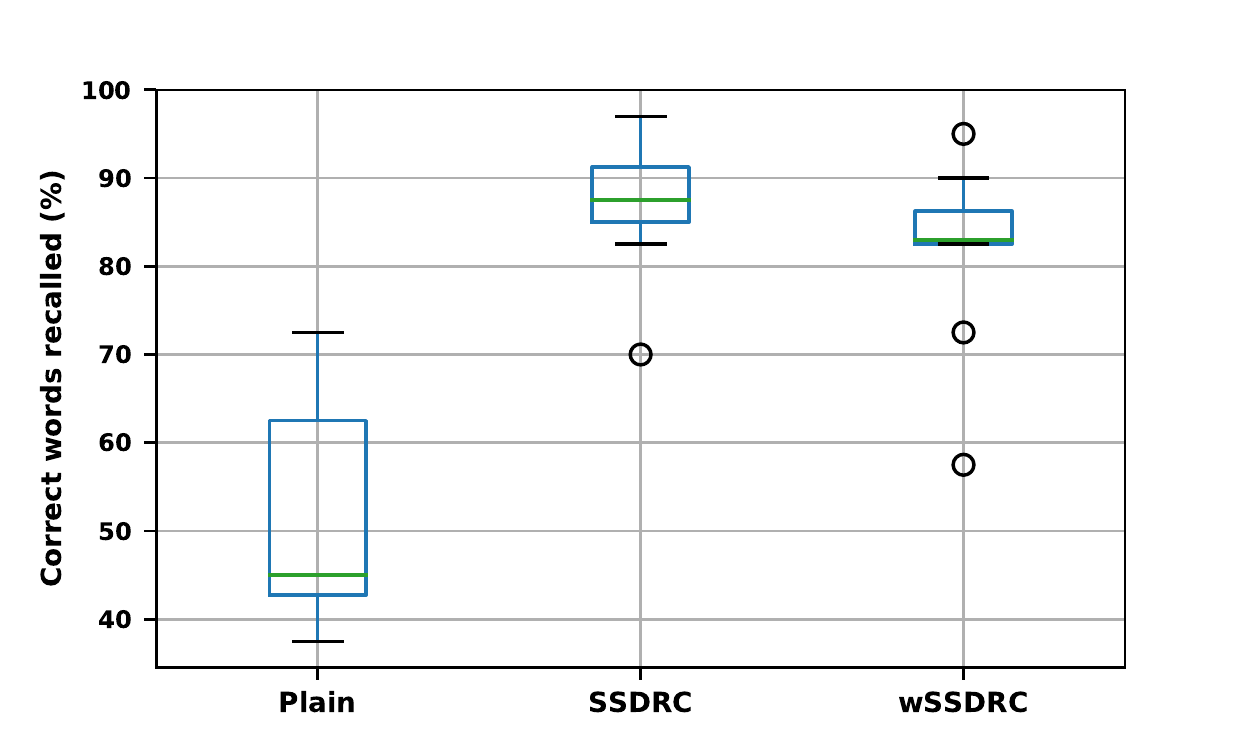}
  \caption{Words recalled by participants Hearing Impairment (HI) in different conditions; boxes represent data dispersion}
  \label{fig:HI}
\end{figure}

The percentage of correct words recalled in each condition from the $13$ participants with normal hearing and $11$ with hearing impairment are plotted in Figures~\ref{fig:NH} and~\ref{fig:HI}, respectively. The median for each condition is represented by the horizontal line inside the box. The variability among the participants' responses is illustrated by the box length: the longer the box, the larger the deviation among participants in that condition. The responses that largely deviate from the rest of the population are encircled as outliers.

 The intelligibility score of plain, unmodified speech for both groups, with NH and HI, is on median 58\% and 45\%, respectively. The values confirm that participants in each group on average listened to the plain test at the SRT points.

Looking at the group with NH, we observe that the neural enrichment model (wSSDRC) has induced a median intelligibility of 97\%, a rise of 39\% from the plain unprocessed speech. SSDRC has produced an median gain of 98\%, a value closely matching that of the wSSDRC model. The difference between the two results is not statistically significant. 

Regarding the group with HI, the median intelligibility of the samples from the neural model (wSSDRC) was 83\% , which is an improvement of 38\% over the Plain condition. SSDRC produced a slightly higher gain of 88\% . This might be due to the few outliers in the wSSDRC condition, as can be seen in Figure~\ref{fig:HI}, which have caused the larger median deviation between SSDRC and wSSDRC, in contrast to the group with NH.

To statistically account for this variability among the groups, and observe its influence on the between group variability, an one-way analysis of variance (ANOVA) has been conducted.

\subsection{Statistical comparison using ANOVA test}
ANOVA is a comparative measure of variance among and between groups. If within-group variability is more significant than between-group variability, the dominance of one group over the other should not be appraised as a reliable gain. ANOVA examines these variations in a more absolute statistical way. In the present study, this is important in order to capture the real gain, if any, as different processing types vs. unprocessed speech are being compared, and more importantly, in order to match the performance of SSDRC with that of wSSDRC, and investigate how close the two models are.

ANOVA computes F-statistics, which is the ratio of inter-group to intra-group variability. Higher F-value indicates higher inter-group variability, which in turn means one group is dominant over the other. The p-value accompanying the F-value indicates that the probability of the predicted F-value could be random. Lower p value indicates higher confidence of the returned F-value.

Firstly, let us consider the NH group. On the null hypothesis that the three modifications -- Plain, SSDRC and wSSDRC -- produce the same intelligibility gain, we ran the one way ANOVA over the three methods. It rendered the result ($F = 163.6, p = 7.4 \times 10^{-18}$), the very high F and very low p indicates that at least one of the compared groups is significantly different. 
Though it is obvious from Figure~\ref{fig:NH} which group falls behind, we have computed an additional series of ANOVA; dividing the three pair groups into sub groups of two pairs. The Plain -- SSDRC  produces ($F = 211.2, p = 9.36 \times 10^{-13}$),  Plain -- wSSDRC produces ($F = 184.5, p = 3.56 \times 10^{-12}$), and SSDRC-wSSDRC produces ($F = 0.192, p = 0.66$).
The picture is clearer now that Plain class is significantly farther from the other two categories. More importantly, when comparing the SSDRC with wSSDRC the F-value is 0.192, which is very close to the ideal case, $F = 0$, the case where the two categories would be exactly equal. This confirms that the wSSDRC produces an equivalent statistical intelligibility gain as the SSDRC for NH.

In the case of the HI group, when performed the statistical test between SSDRC -- Plain categories, the statistics shows ($F = 65.3, p = 1.02 X 10^{-7}$), while the neural enrichment (wSSDRC) -- Plain gives  ($F = 39.28, p = 4.04 X 10^{-6}$). Though the F-values are not as large as the NH, here also, the higher F values indicate the obvious fact that the processing has resulted in substantial intelligibility gain.
Though the two F values differ significantly, when computing the same test between SSDRC -- wSSDRC the F score  ($F = 1.94, p = 0.178$) was close to the matching point,  which again manifests that both models are rendering relatively similar gain. 

\vspace{.2cm}
The ANOVA tests further confirm the fact that the neural enrichment model (wSSDRC) produces an equivalent intelligibility gain with the signal processing model (SSDRC) that was used to train the model. As a whole, the study confirms that a carefully designed neural model could learn the speech modification task even on a language like Greek which differs from languages of Latin origin. Though it may not be attractive at this point, the same neural model could have been robust against noise if it were trained with noise perturbations as input, contrasting to the signal processing model. This is still work in progress and further results will be reported in the future. As such, the finding that neurally modified samples are equally intelligible supports the future of neural enrichment models.
Few samples from the wSSDRC model are displayed here \footnote{\url{https://www.csd.uoc.gr/~shifaspv/IS2020-demo}},
and a tensorflow implementation of the model is provided \footnote{\url{https://github.com/shifaspv/wSSDRC-tesnorflow-implementation}}.

\section{Conclusion}
\label{Section.6}

In this paper, we presented the results of a subjective evaluation of a neural speech enrichment model for Greek language. The neural model was trained to mimic the intelligibility modification of a standard, well-known signal processing approach called SSDRC. The recently created Greek Harvard corpus was used for training and evaluation of the model. An extensive subjective evaluation has been carried out with listeners with normal hearing and hearing impairment. It is shown that the samples generated by the neural model are well more intelligible than the plain unprocessed speech - the intuition has been confirmed with the one-way ANOVA statistical test.
When compared to the signal processing approach, the neural enrichment model produced an equivalent intelligibility boost for both the listening groups. The findings confirm that an effectively designed neural model could learn and generalize the speech modification task aiming at intelligibility improvement. 
Besides, the neural enrichment model has the inherent advantage of being noise resistive, and would be a replacement for the signal processing approach in conditions where noise adversities are expected.

\section{Acknowledgements}
The authors would like to thank Dr. George P. Kafentzis (Post-doctoral Researcher, Adjunct Lecturer, Computer Science Department, University of Crete, Greece) for his kind help on conducting the listening test and organizing the manuscript.\\
This work was funded by the E.U. Horizon2020 Grant
Agreement 675324, Marie Sklodowska-Curie Innovative Training Network, ENRICH.

\bibliographystyle{IEEEbib}
\bibliography{strings,refs}

\end{document}